\documentclass[english,twocolumn,rmp]{revtex4}
\usepackage[T1]{fontenc}
\usepackage[latin9]{inputenc}
\usepackage{amstext}
\usepackage{amssymb}
\usepackage{graphicx}
\usepackage{esint}

\makeatletter
\@ifundefined{textcolor}{}
{%
 \definecolor{BLACK}{gray}{0}
 \definecolor{WHITE}{gray}{1}
 \definecolor{RED}{rgb}{1,0,0}
 \definecolor{GREEN}{rgb}{0,1,0}
 \definecolor{BLUE}{rgb}{0,0,1}
 \definecolor{CYAN}{cmyk}{1,0,0,0}
 \definecolor{MAGENTA}{cmyk}{0,1,0,0}
 \definecolor{YELLOW}{cmyk}{0,0,1,0}
}

\makeatother

\usepackage{babel}
\begin{document}

\title{Response to Selection for continuous traits: an alternative to the
breeder's and Lande's equations.}

\author{Bahram Houchmandzadeh }

\address{CNRS/Univ. Grenoble 1, LIPhy UMR 5588, Grenoble, F-38041, France
.}
\begin{abstract}
The breeder's equation is a cornerstone of quantitative genetics and
is widely used in evolutionary modeling. The equation, which reads
$R=h^{2}S$, relates response to selection $R$ (the mean phenotype
of the progeny) to the selection differential $S$ (mean phenotype
of selected parents) through a simple proportionality relation. The
validity of this relation however relies strongly on the \emph{normal}
(Gaussian) distribution of the parent genotype, which is an unobservable
quantity and cannot be ascertained. In contrast, we show here that
if the fitness (or selection) function is Gaussian, an alternative,
exact linear equation of the form $R'=j^{2}S'$ can be derived, regardless
of the \emph{parental genotype} distribution. Here $R'$ and $S'$
stand for the mean phenotypic \emph{lag} with respect to the mean
of the fitness function in the offspring and selected populations.
To demonstrate this relation, we derive the exact functional relation
between the mean phenotype in the selected and the offspring population
and deduce all cases that lead to a linear relation between the mean
phenotypes of progeny and selected parents. These results, which are
confirmed by individual based numerical simulations, generalize naturally
to the concept of $G$ matrix and the multivariate Lande's equation
$\Delta\mathbf{\bar{z}}=GP^{-1}\mathbf{S}$. The linearity coefficients
of the alternative equation are not changed by selection. The alternative
equation can thus be more suitable for long term evolutionary studies
than the $G$ matrix. 
\end{abstract}

\keywords{%
\thanks{response to selection, additive genetic effects, quantitative genetics.%
}}

\maketitle

\section{Introduction.}

The breeder's equation for the evolution of quantitative traits for
additive genetic effects, introduced by Lush \cite{Lush1943} is widely
used both in artificial and natural selection theory and experiments
\cite{Falconer1995,Lynch1998,LANDE1976,Heywood2005} and appears in
all textbooks of quantitative genetic. The scalar equation $R=h^{2}S$,
or its vectorial version $\Delta\bar{\mathbf{z}}=GP^{-1}\mathbf{S}$
ascertain that the response to selection (mean phenotype of offspring)
and the selection differential (mean phenotype of selected parents)
are related through a linear relation which is the ratio of genotype
to phenotype variances, $h^{2}$. 

Use of the breeder's equation and its underlying assumptions have
been criticized by many authors \cite{Heywood2005,Gienapp2008,Pigliucci2006,Kruuk2004,Pemberton2010}.
One fundamental assumption of the breeder's equation is the normal
(Gaussian) distribution of the breeding value (genotype) \emph{and}
environment factors. Authors who demonstrate the linear relation \cite{Falconer1995,Lynch1998,Kimura1978,Crow2009,LANDE1979,LANDE1983,Nagylaki1992}
assume normal distribution for the above quantities or the analogous
hypothesis of linearity of the parent-offspring regression (see Appendix/Parent-offspring
regression). When this assumption is relaxed, the breeder's equation
is no longer valid and one has to resort to a system of hierarchical
moment (or alternatively, cumulant) equations to describe the changes
; in general, this system is not closed and the moments of a given
order  depend on moments of higher order \cite{TURELLI1990}. 

The assumption of a Gaussian distribution of the genotype can be criticized
on several grounds \cite{Pigliucci2006,Pigliucci1997,Geyer2008}.
For example, the very act of selection causes the genotype distribution
to deviate from a Gaussian \cite{TURELLI1990,TURELLI1994} (see also
equation \ref{eq:genotyperecurrence} below). Another important case
is when the genotype is a cross between different breeds due to external
gene flow or the breeder's scheme. In many cases, the phenotype can
have a bell shape and thus is assumed to be Gaussian, when the genotype
is indeed far from it (see for example figure 2a). It is sometimes
argued that even if the breeding value does not follow a normal distribution,
a scale can be used to restore it to a normal distribution. Such a
scale however will also distort the distribution of environment factors
and the assumptions of breeder's equation are violated even in this
case. 

For \emph{additive} genetic effects and in the absence of epistasis
and dominance, I derive here a precise functional relation between
the mean of the trait in the selected subpopulation and in their progeny
for the  general case. The mathematical formulation is close to the
framework used by many authors such as Slatkin, Lande and Karlin \cite{Slatkin1970,LANDE1979,KARLIN1979}.
I then use a standard tool of functional analysis, the Fourier transform,
to deduce all the cases taht lead to a linear relation between the
response $R$ and the selection differential $S$, regardless of the
selection function. These cases imply a precise form of the distributions
of genotype and environment factors, and I show that the proportionality
factor between $R$ and $S$ is the heritability coefficient $h^{2}$
only if these distributions are \emph{normal}. 
\begin{figure}
\begin{centering}
\includegraphics[width=0.8\columnwidth]{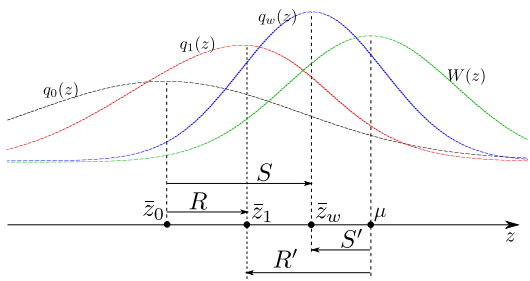}
\par\end{centering}

\caption{Schematic representation of the selection lag $S'$, the response
lag $R'$ and their relation to the selection differential $S$ and
the response $R$. The mean phenotype of parental generation $\bar{z}_{0}$,
selected population $\bar{z}_{w}$, the progeny $\bar{z}_{1}$ and
the peak of selection function$\mu$ are represented on the phenotype
axis $z$. Dashed curves represent a sketch of the distributions of
parental phenotype $q_{0}(z)$, selected parents $q_{w}(z)$, the
progeny $q_{1}(z)$ and the selection function $W(z)$. \label{fig:fig1-1}}

\end{figure}

The genotype however is not observable or controllable and its normal
distribution cannot be assumed a priori. I show that if instead of
the genotype, the \emph{fitness function} and environment factors
are Gaussian, then a new linear relation can be obtained in the form
of 
\begin{equation}
R'=j^{2}S'\label{eq:altbreed}
\end{equation}
\emph{regardless} of the genotype distribution. Here $R'$ and $S'$
are the mean phenotypic lag with respect to the mean of the fitness
function of the progeny and the selected population (Figure 1). The
$j^{2}$ coefficient contains only the width of the fitness function
and environment factors. The use of a Gaussian selection function,
both in artificial and natural selection (as an approximation of stabilizing
selection) is widespread \cite{LANDE1976,Kimura1978,LEWONTIN1964,Zhang2010}
and the above relation is potentially as useful as the standard breeder's
equation. 

The advantage is more critical when the breeder's or Lande's equations
are used in long term evolution, where the variance of the genotype
(or the $G$ matrix) also varies and $h^{2}$ cannot be assumed to
remain constant \cite{Pigliucci1997,Gavrilets1995,Roff2000} ; in
contrast, the relation (\ref{eq:altbreed}) remains valid if each
round of selection uses a Gaussian fitness function. 

The above results generalize naturally to multivariate trait selection
where the alternative Lande's equation is 
\begin{equation}
\mathbf{R}'=(\Omega+E)\Omega^{-1}\mathbf{S}'\label{eq:altland}
\end{equation}
where $\mathbf{R}'$ and $\mathbf{S}'$ are the vectorial phenotype
lag, and $\Omega$ and $E$ are the covariance matrices of the fitness
function and the environment respectively.

The Fisher's fundamental theorem states that ``the rate of increase
in fitness of any organism at any time is equal to its genetic variance
in fitness at that time''. The alternative equations (\ref{eq:altbreed})
or (\ref{eq:altland}) could thus seem unusual, as the linearity coefficient
or matrix does not contain the genetic variance. There is however
no contradiction : both quantities $R'$ and $S'$ depend on the genetic
variance but their ratio is not. All the above results are confirmed
by individual based numerical simulations. 

This article is organized as follows : in the Results section, I first
derive the general functional relationship between $R$ and $S$;
the second subsection is devoted to all the cases where these two
quantities can be linearly related, including the special case of
the breeder's equation. The alternative breeder's equation is derived
in the third subsection and all the results are generalized to selection
on multiple traits in the fourth subsection. The above results are
put into perspective in the Discussion section. Technical details
such as the use of Fourier transforms and numerical simulations are
treated in the Appendix.

\section{Results.}

\subsection*{General results.}

Consider a continuous phenotype $Z$, which is the result of additive
genetic effect $Y$ and the environment $\xi$ \cite{Visscher2008}
\[
Z=Y+\xi
\]
The term \emph{environment} encompasses here any source of \emph{noise}
that causes the observed phenotype $z$ to deviate from the (unobserved)
breeding value $y$ \cite{Wright1920,Lynch1998,Raj2008}. In the following,
the population distribution of the breeding value (genotype) and its
variance in the parental generation are denoted $p_{0}(y)$ and $\sigma_{A}^{2}$.
The environment effect is captured by the distribution law $f(z|y)$,
the probability density of observing phenotype $z$ with the given
genotype $y$. We will suppose that $f$ is a symmetric function of
its argument of the form $f(z|y)=f(z-y)$. 

A subpopulation among the parental generation is selected according
to a fitness or selection function $W(z)$, the proportion of phenotypes
in $[z,z+dz[$ to be selected for the production of the next generation.
The selected individuals produce offspring which will constitute the
next generation. As we will show below, the response $R$ (the mean
of the phenotype trait in the offspring) and the selection differential
 $S$ (the mean of the phenotype trait in the selected parents) are
given by 
\begin{eqnarray}
R=E(Z_{1}) & = & \frac{1}{\bar{W}}\iint_{\mathbb{R}^{2}}yp_{0}(y)W(z)f(z-y)dydz\label{eq:genresponse}\\
S=E(Z_{w}) & = & \frac{1}{\bar{W}}\iint_{\mathbb{R}^{2}}zp_{0}(y)W(z)f(z-y)dydz\label{eq:gensel}
\end{eqnarray}
where $\bar{W}$ is the mean fitness of parental generation. The above
equations (\ref{eq:genresponse},\ref{eq:gensel}) are used for example
by Lande \cite{LANDE1979}, although their derivation there depended
on the normal distribution of the genotype. I derive these equations
here for the more general case. 

Before going into the details of calculations, note that the genotype
distribution $p_{0}(y)$ and the selection function $W(z)$ play a
symmetric role in the above expressions. In the following sections,
we will explore specific functional forms of $p_{0}(y)$ and $W(z)$
which lead to a linear relationship between $R$ and $S$. Because
of the symmetric role of these two functions however, once a particular
relation is obtained for a specific form of $p_{0}(y)$ regardless
of $W(z)$, an analogous relationship can be obtained for a similar
form of $W(z)$ regardless of $p_{0}(y)$. This is what leads us to
an alternative form of the breeder's equation. 

Let us now derive the equations (\ref{eq:genresponse},\ref{eq:gensel}).
We note that the distribution of the phenotype $Z$ in the parental
generation is given by 
\begin{equation}
q_{0}(z)=\int_{\mathbb{R}}p_{0}(y)f(z|y)dy\label{eq:q0}
\end{equation}
We will denote its variance by $\sigma_{P}^{2}$. 

The distribution of the phenotype $z$ in the parental population
selected according to the fitness function $W(z)$ is 
\[
q_{w}(z)=\frac{1}{\bar{W}}q_{0}(z)W(z)
\]
where $\bar{W}$ is the mean fitness of the parental generation 
\begin{eqnarray*}
\bar{W} & = & \int_{\mathbb{R}}q_{0}(z)W(z)dz\\
 & = & \iint_{\mathbb{R}^{2}}p_{0}(y)W(z)f(z|y)dydz
\end{eqnarray*}
The genotype distribution of the selected population is \cite{TURELLI1994}
\begin{eqnarray}
p_{w}(y) & = & \frac{1}{\bar{W}^{\dagger}}\int_{\mathbb{R}}p_{0}(y)f(z|y)W(z)dz\label{eq:genotyperecurrence}\\
 & = & \frac{1}{\bar{W}^{\dagger}}p_{0}(y)W^{\dagger}(y)\label{eq:pselectedparents}
\end{eqnarray}
where 
\begin{equation}
W^{\dagger}(y)=\int_{\mathbb{R}}W(z)f(z|y)dz\label{eq:Wtilde}
\end{equation}
is the \emph{genotype} fitness function, \emph{i.e. }the convolution
of the phenotype fitness function by the environment factors. $\bar{W}^{\dagger}$
is the mean genotype fitness :
\begin{eqnarray*}
\bar{W}^{\dagger} & = & \int_{\mathbb{R}}p_{0}(y)W^{\dagger}(y)dy\\
 & = & \iint_{\mathbb{R}^{2}}p_{0}(y)W(z)f(z|y)dydz
\end{eqnarray*}
 Note that $\bar{W}=\bar{W}^{\dagger}$ as both these quantities are
defined by the same double integration over the domains of $y$ and
$z$. 

For a large, randomly mating population, reproduction gives for the
distribution of breeding values in the next generation \cite{Slatkin1970,KARLIN1979,Bulmer1985,TURELLI1994}
\[
p_{1}(y)=\iint_{\mathbb{R}^{2}}p_{w}(y_{a})p_{w}(y_{b})L\left(y-(y_{a}+y_{b})/2\right)dy_{a}dy_{b}
\]
The exact form of the probability density $L(y)$ that captures the
inheritance process (recombination, segregation, ...) is not important
here ; Turelli and Barton (\cite{TURELLI1994}) for example use a
normal distribution for $L(y)$ in the framework of the infinitesimal
model. For our purpose, it is enough to suppose that the mean of the
distribution $L(y)$ is zero, \emph{i.e. }$\int_{y}yL(y)dy=0$ which
is valid in the absence of dominance and epistasis effects \cite{TURELLI1990}
(see also Appendix/Segregation density function). 

The phenotype distribution of the progeny is 
\begin{equation}
q_{1}(z)=\int_{\mathbb{R}}p_{1}(y)f(z|y)dy\label{eq:q1}
\end{equation}
We now make the further assumption that (i) the environment and genotype
are independent random variables, so that $f(z|y)=f(z-y)$ and therefore
the variances are additive : $\sigma_{P}^{2}=\sigma_{A}^{2}+\sigma_{E}^{2}$
and (ii) environment effects are of zero mean ($\int_{x}xf(x)dx=0$)
and symmetric ($f(-x)=f(x)$ ). An environmental noise with such a
distribution law does not change the mean of the random variable :
$E(Z)=E(Y+\xi)=E(Y)$. Therefore, the mean phenotype of the offspring
is 
\begin{eqnarray}
R=E(Z_{1}) & = & E(Y_{1})\nonumber \\
 & = & \int_{\mathbb{R}}yp_{1}(y)dy\nonumber \\
 & = & (1/2)\iint_{\mathbb{R}^{2}}(y_{a}+y_{b})p_{w}(y_{a})p_{w}(y_{b})dy_{a}dy_{b}\nonumber \\
 & = & \int_{\mathbb{R}}yp_{w}(y)dy\label{eq:Rinfinitesimal}\\
 & = & \frac{1}{\bar{W}}\iint_{\mathbb{R}^{2}}yp_{0}(y)W(z)f(z-y)dydz\label{eq:Rfinal}
\end{eqnarray}
which is the equation (\ref{eq:genresponse}). Note that the first
lines of the above equations merely state that the \emph{expectations}
of the breeding's value of parent and offspring are equal for purely
additive traits.

On the other hand, the mean phenotype of the selected parents is 
\begin{eqnarray*}
S=E(Z_{w}) & = & \int_{\mathbb{R}}zq_{w}(z)dz\\
 & = & \frac{1}{\bar{W}}\int_{\mathbb{R}}zq_{0}(z)W(z)dz\\
 & = & \frac{1}{\bar{W}}\iint_{\mathbb{R}^{2}}zp_{0}(y)W(z)f(z-y)dydz
\end{eqnarray*}
which is the equation (\ref{eq:gensel}). 

Note that for an asexually reproducing organism, or for a sexually
reproducing population which remains at Hardy-Weinberg equilibrium
after selection-reproduction, we would have $p_{1}(y)=p_{w}(y)$ ;
this would again lead to the same equation (\ref{eq:Rinfinitesimal})
and the same response (\ref{eq:genresponse}). The conditions for
the existence of multilocus Hardy-Weinberg equilibrium were analyzed
by Karlin and Liberman \cite{KARLIN1979a,KARLIN1979b} who concluded
that for additive traits, the equilibrium is stable for a wide range
of recombination distributions.

\subsection*{Conditions for proportionality of $R$ and $S$.}

The relations (\ref{eq:genresponse},\ref{eq:gensel}) show that the
selection differential $S$ and the response $R$ to it are related
through a functional equation involving three factors : genotype distribution,
the selection function and the environmental noise. It is far from
obvious that $R$ and $S$ could be proportional. 

Fourier transforms (FT) in functional analysis play a role analogous
to logarithms in algebra, and part of their usefulness is due to the
fact that they transform convolution products into simple products.
They are useful for clarifying the $R-S$ relation, where we can transform
the double integrations into simple ones. Here $\tilde{u}(k)$ designates
the FT of the function $u(x)$ and $a*$ is the complex conjugate
of $a$ (see Appendix/Fourier Transforms). We set the origin of the
breeding values at its mean in the parental population, \emph{i.e.
$\int_{\mathbb{R}}yp_{0}(y)dy=0$}. The response and selection differential
are 
\[
R=\frac{i}{2\pi\bar{W}}\int_{\mathbb{R}}\tilde{W}^{*}(k)\tilde{p}_{0}'(k)\tilde{f}(k)dk
\]
and 
\begin{eqnarray}
S & = & \frac{i}{2\pi\bar{W}}\int_{\mathbb{R}}\tilde{W}^{*}(k)\left[\tilde{p}_{0}'(k)\tilde{f}(k)+\tilde{p}_{0}(k)\tilde{f}'(k)\right]dk\nonumber \\
 & = & R+\frac{i}{2\pi\bar{W}}\int_{\mathbb{R}}\tilde{W}^{*}(k)\tilde{p}_{0}(k)\tilde{f}'(k)dk\label{eq:FTsel}
\end{eqnarray}
We see that $S$ and $R$ can be proportional if the second term of
the r.h.s. of equation (\ref{eq:FTsel}) is proportional to $R$ ;
this will be true, \emph{regardless} of the selection function $W$,
if 
\begin{equation}
\tilde{p}_{0}(k)\tilde{f}'(k)=a\tilde{p}_{0}'(k)\tilde{f}(k)\label{eq:propcondition}
\end{equation}
where $a$ is an arbitrary constant. Equation (\ref{eq:propcondition})
is the necessary and sufficient condition that defines the functional
shape of the genotype distribution and the environment noise compatible
with the proportionality of $R$ and $S$ regardless of the selection
function. If condition (\ref{eq:propcondition}) is fulfilled, then
\[
R=(1+a)^{-1}S
\]
On the other hand, eq. (\ref{eq:propcondition}) can be seen as a
differential equation whose solution is given by 
\begin{equation}
\tilde{f}(k)=b\tilde{p}_{0}(k)^{a}\label{eq:propcondi2}
\end{equation}
where $b$ is another arbitrary constant. 

If $\tilde{f}(k)$ and $\tilde{p}_{0}(k)$ are both Gaussians, \emph{i.e.,
}
\begin{eqnarray*}
\tilde{f}(k) & = & \text{\ensuremath{\exp}}\left(-\sigma_{E}^{2}k^{2}/2\right)\\
\tilde{p}_{0}(k) & = & \exp\left(-\sigma_{A}^{2}k^{2}/2\right)
\end{eqnarray*}
then the relation(\ref{eq:propcondi2}) is satisfied by 
\[
a=\sigma_{E}^{2}/\sigma_{A}^{2}
\]
and we retrieve the usual breeder's equation $R=h^{2}S$ where $h^{2}=\sigma_{A}^{2}/(\sigma_{A}^{2}+\sigma_{E}^{2})$.
Of course, if $\tilde{f}(k)$ and $\tilde{p}_{0}(k)$ are of the above
form, their inverse Fourier transforms represent normal distributions
of width $\sigma_{E}$ and $\sigma_{A}$ respectively (see Appendix/Fourier
Transforms). 

We see however that even if the strict condition (\ref{eq:propcondi2})
is fulfilled, the proportionality constant need not be $h^{2}$. Consider
for example the class of stretched exponential functions $\phi(k)=\exp(-|k|^{\alpha})$
which generalizes Gaussians (case $\alpha=2$). Set $\tilde{f}(k)=\phi(\sigma_{E}k)$,
$\tilde{p}_{0}(k)=\phi(\sigma_{A}k)$. The inverse Fourier transform
of these functions gives the distribution of the genotype $Y$ and
environment effect $E$ and it is straightforward to show that as
for the Gaussian case, $\mbox{Var}(E)/\mbox{Var}(Y)=\sigma_{E}^{2}/\sigma_{A}^{2}$.
Condition (\ref{eq:propcondi2}) however is satisfied this time with
$a=\sigma_{E}^{\alpha}/\sigma_{A}^{\alpha}$ and therefore the realized
heritability $h^{\alpha}=R/S$ is 
\[
h^{\alpha}=\frac{\sigma_{A}^{\alpha}}{\sigma_{A}^{\alpha}+\sigma_{E}^{\alpha}}
\]
The above examples were to emphasize the fact that selection-independent
proportionality is achieved only for particular pairs of genotype/environment
distributions. In general, as shown in figure 2, the realized heritability
is not constant and depends critically on the selection function $W(z)$.
\begin{figure}
\begin{centering}
\includegraphics[width=0.8\columnwidth]{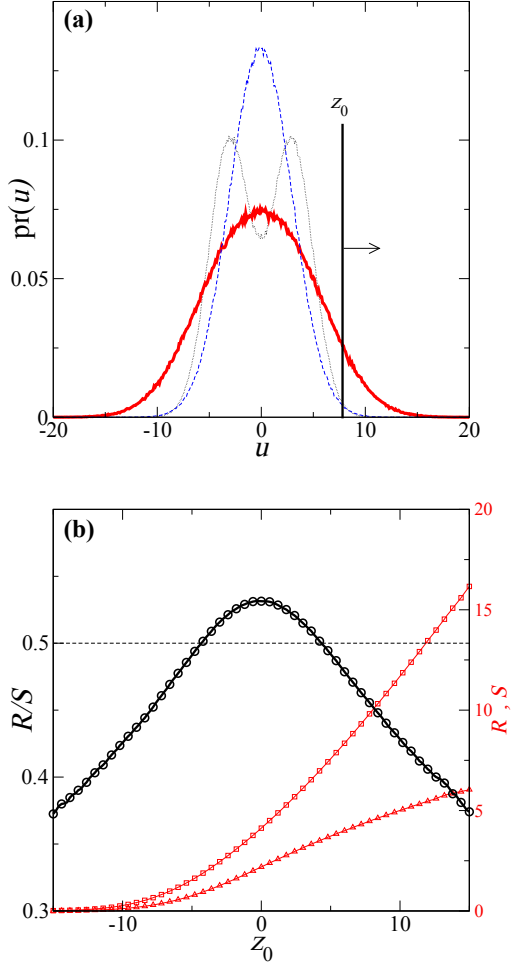}
\par\end{centering}

\caption{Individual based numerical simulation of the selection process and
its comparison to theoretical results. (a) $10^{6}$ parental individuals
are generated. The breeding value $y_{0,i}$ of an individual is drawn
from a double Gaussian distribution ; environmental effects $e_{i}$
are drawn from a normal distribution with the same variance ; the
phenotype $z_{i}$ of an individual is determined by adding the environment
effect to its breeding value $z_{0,i}=y_{0,i}+e_{i}$. Distributions
of $p_{0}(y)$ (dotted curve, black), $f(x)$ (dashed curve, blue)
and $q_{0}(z)$ (thick solid line, red) are shown. A truncation selection
for high phenotype $z>z_{0}$ is applied and shown as the thick vertical
line. (b) selected individuals produce progeny with the same breeding
value $(y_{1,i}=y_{w,i})$ which are subject to the same random environmental
effect : $z_{1,i}=y_{1,i}+e'_{i}$. The mean phenotype is computed
from the individuals in each generation and is a function of truncation
threshold $z_{0}$. Right scale: $R=E(z_{1})$ (triangles, red) and
$S=E(z_{w})$ (squares, red) as a function of $z_{0}$. Left scale
: the realized heritability $R/S$ (circles, black) as a function
of $z_{0}$ ; the value of $h^{2}=\mbox{Var}(Y_{0})/\mbox{Var}(Z_{0})$
is indicated by the horizontal dashed line (black).}

\end{figure}

\subsection*{Alternative breeder's equation.\label{sec:Alternative-breeder's-equation.}}

Optimal phenotypic selection approximated by Gaussians has been considered
by many authors both in artificial (as early as Lush \cite{Lush1943}
) and in natural selection (as early as Wright\cite{Wright1935} Haldane\cite{Haldane1954})
and it is widespread in the literature \cite{LEWONTIN1964,LANDE1976,Kimura1978,KARLIN1979a,Zhang2010}.
If the selection function is Gaussian, a new linear relation can be
extracted from the general relations (\ref{eq:genresponse},\ref{eq:gensel}),
\emph{regardless} of the (unobservable) breeding value distribution.

Note that a symmetric role is played by $W(z)$ and $p_{0}(y)$ in
the general expressions (\ref{eq:genresponse},\ref{eq:gensel}).
Hence permuting their role will lead us, following the same line of
arguments, to deduce all linear cases regardless of genotype. Equations
(\ref{eq:genresponse},\ref{eq:gensel}) are obtained by multiplying
the function $F(y,z)=W(z)p_{0}(y)f(z-y)$ either by $y$ or $z$ and
integrating over $\mathbb{R}^{2}$. In order to obtain the breeder's
equation of the previous section, we wrote the integration over the
$y$ variable as a convolution product and performed the Fourier Transform
on the $z$ variable. 

On the other hand, we could have proceeded by writing eqs. (\ref{eq:genresponse},\ref{eq:gensel})
first as a convolution product on $z$ and then perform a Fourier
transform on the variable $y$ (see Appendix/Fourier Transform). In
this case, we get 
\begin{equation}
S=\frac{i}{2\pi\bar{W}}\int_{\mathbb{R}}\tilde{p}^{*}(k)\tilde{W}'(k)\tilde{f}(k)dk\label{eq:altsel}
\end{equation}
 and 
\begin{eqnarray}
R & = & \frac{i}{2\pi\bar{W}}\int_{\mathbb{R}}\tilde{p}^{*}(k)\frac{d}{dk}\left(\tilde{W}(k)\tilde{f}(k)\right)dk\label{eq:altres}
\end{eqnarray}
The arguments of the previous section can be repeated. Let us center
the selection function by setting $W(z)=W_{c}(z-\mu)$ where 
\[
\mu=\int_{\mathbb{R}}zW(z)dz
\]
Then 
\begin{equation}
S'=(S-\mu)=\frac{i}{2\pi\bar{W}}\int_{\mathbb{R}}\tilde{p}^{*}(k)e^{-ik\mu}\tilde{W_{c}}'(k)\tilde{f}(k)dk\label{eq:Sprime}
\end{equation}
and 
\begin{equation}
R'=(R-\mu)=\frac{i}{2\pi\bar{W}}\int_{\mathbb{R}}\tilde{p}^{*}(k)e^{-ik\mu}\frac{d}{dk}\left(\tilde{W_{c}}(k)\tilde{f}(k)\right)dk\label{eq:Rprime}
\end{equation}
The quantities $S'$ and $R'$ are alternate selection differential
and response and represent the \emph{lag} with respect to the mean
of the selection function (figure 1). In the case where the selection
function and the environment factors are both normally distributed
with width $\sigma_{W}$ and $\sigma_{E}$, a repetition of the arguments
of the previous sections leads to 
\begin{equation}
R'=j^{2}S'\label{eq:alt_breeder}
\end{equation}
 where 
\[
j^{2}=\frac{\sigma_{W}^{2}+\sigma_{E}^{2}}{\sigma_{W}^{2}}
\]
We stress that relation (\ref{eq:alt_breeder}) is obtained \emph{regardless}
of the unknown genotype distribution $p_{0}(y)$. Figure 3 illustrates
the accuracy of this new relation compared to the usual breeder's
equation. As noted by Turelli and Barton \cite{TURELLI1990,TURELLI1994},
the discrepancy in the standard breeder's equation predictions is
highest for weak selection. 
\begin{figure}
\begin{centering}
\includegraphics[width=0.8\columnwidth]{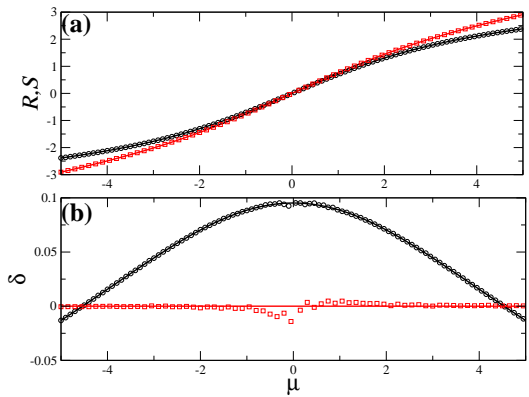}
\par\end{centering}

\caption{Individual based numerical simulation of Gaussian selection and the
response to it as described in figure 2 ; The selection function here
is a Gaussian $\exp\left((z-\mu)^{2}/2\sigma_{W}^{2}\right)$. (a)
The mean phenotype of the selected $S=E(Z_{w})$ (red squares) and
the progeny $R=E(Z_{1})$ (black circles) as a function of the optimum
phenotype $\mu$. (b) the \emph{deviation} $\delta$ from their theoretical
values of the of the realized heritability $R/S-h^{2}$ (black circles)
and and its alternative form $R'/S'-j^{2}$ (red squares) as a function
of $\mu$. Continuous traits represent theoretical values obtained
from eqs (\ref{eq:genresponse},\ref{eq:gensel})}

\end{figure}

If the selection function, and the distributions of genotype and environment
factors are all Gaussian functions, the standard and alternative breeder's
equation can be combined, which leads to a simple linear relation
\begin{eqnarray}
S & = & \alpha\mu\label{eq:allgauss1}\\
R & = & \alpha h^{2}\mu\label{eq:allgauss2}
\end{eqnarray}
where $\alpha=(j^{2}-1)/(j^{2}-h^{2})$. Relations (\ref{eq:allgauss1},\ref{eq:allgauss2})
can be used as a test for normal distribution of the genotype. 

The alternative equation (\ref{eq:alt_breeder}) is not in contradiction
with Fisher's fundamental theorem and does not predict evolution independently
of genetic variance. Both $R'$ and $S'$ are dependent on the genetic
variance, as can be seen in the general equations (\ref{eq:genresponse}-\ref{eq:gensel})
; the coefficient of the linear equation (\ref{eq:alt_breeder}) relating
them however is free of genetic variance. Consider for example the
extreme case where there is no genetic variance ($\sigma_{A}=0$).
The distribution of the breeding value then becomes a Dirac's delta
function $p_{0}(y)=\delta(y)$ and the value of $R$ and $S$ are
readily obtained from equations (\ref{eq:genresponse}-\ref{eq:gensel}):
\begin{eqnarray}
R & = & 0\label{eq:z1}\\
S & = & \mu\frac{\sigma_{E}^{2}}{\sigma_{W}^{2}+\sigma_{E}^{2}}\label{eq:zw}
\end{eqnarray}
Therefor $R'=-\mu$, $S'=-\mu/j^{2}$ and equation (\ref{eq:alt_breeder})
is verified trivially.

\subsection*{Selection on multiple traits.}

The results of the above sections are naturally generalized to selection
on multiple traits. Consider the vectors of parental breeding values
$\mathbf{y}_{0}=(y_{1},y_{2},...y_{N})$, environmental effects $\mathbf{e}=(e_{1},...,e_{N})$
and their phenotype $\mathbf{z}_{0}=\mathbf{y}_{0}+\mathbf{e}$ ,
to which a selection function $W(\mathbf{z})$ is applied. Using the
same notations as in the previous sections, we find without difficulty
that 
\begin{eqnarray*}
\bar{\mathbf{z}}_{1} & = & \int_{\mathbb{R}^{N}\times\mathbb{R}^{N}}\mathbf{y}p_{0}(\mathbf{y})W(\mathbf{z})f(\mathbf{z}-\mathbf{y})d\mathbf{y}d\mathbf{z}\\
\bar{\mathbf{z}}_{w} & = & \int_{\mathbb{R}^{N}\times\mathbb{R}^{N}}\mathbf{z}p_{0}(\mathbf{y})W(\mathbf{z})f(\mathbf{z}-\mathbf{y})d\mathbf{y}d\mathbf{z}
\end{eqnarray*}
As before, using Fourier Transforms, these relations transform into
\begin{eqnarray*}
\bar{\mathbf{z}}_{1} & = & \frac{i}{2\pi}\int_{\mathbb{R}^{N}}\tilde{W}^{*}(\mathbf{k})\left(\nabla\tilde{p}_{0}(\mathbf{k})\right)\tilde{f}(\mathbf{k})d^{M}\mathbf{k}\\
\bar{\mathbf{z}}_{w} & = & \frac{i}{2\pi}\int_{\mathbb{R}^{N}}\tilde{W}^{*}(\mathbf{k})\nabla\left(\tilde{p}_{0}(\mathbf{k})\tilde{f}(\mathbf{k})\right)d^{M}\mathbf{k}
\end{eqnarray*}
where $\nabla$ is the gradient operator: $\nabla f=(\partial f/\partial x_{1},...\partial f/\partial x_{N})$.
We see again that $\mathbf{\bar{z}}_{1}$ and $\mathbf{\bar{z}}_{w}$
are linearly related if 
\[
\tilde{p}_{0}(\mathbf{k})\left(\nabla\tilde{f}(\mathbf{k})\right)=A\left(\nabla\tilde{p}_{0}(\mathbf{k})\right)\tilde{f}(\mathbf{k})
\]
where $A$ is a constant matrix. The linear relation is automatically
satisfied if both $p_{0}$ and $f$ follow a Gaussian distribution
\begin{eqnarray*}
p_{0}(\mathbf{y}) & \propto & \exp\left(-\frac{1}{2}\mathbf{y}^{T}G^{-1}\mathbf{y}\right)\\
f(\mathbf{x}) & \propto & \exp\left(-\frac{1}{2}\mathbf{x}^{T}E^{-1}\mathbf{x}\right)
\end{eqnarray*}
where $G$ and $E$ are the covariance matrices for the genotype and
environmental effects. Defining $P=G+E$ as the phenotype covariance
matrix, it is straightforward to show that in this case $A=EG^{-1}$
and therefore \cite{LANDE1979} 
\[
\bar{\mathbf{z}}_{1}=GP^{-1}\mathbf{\bar{z}}_{w}
\]
which is the usual breeder's equation for multiple traits. We stress
that the limitation of this relation is the same as that of the scalar
version : it relies on the normal distribution of the genotype. On
the other hand, if the selection function $W(\mathbf{z})$ is Gaussian
\[
W(\mathbf{z})\propto\exp\left(-\frac{1}{2}(\mathbf{z}-\mathbf{\mu)}^{T}\Omega^{-1}(\mathbf{z-}\mu)\right)
\]
the arguments of the previous section \ref{sec:Alternative-breeder's-equation.}
can be repeated and lead to the generalization of the alternative
vectorial breeder's equation (\ref{eq:alt_breeder}) 
\[
\bar{\mathbf{z}}_{1}-\mathbf{\mu}=(\Omega+E)\Omega^{-1}(\mathbf{\bar{z}}_{w}-\mathbf{\mu})
\]
which, in analogy with equation (\ref{eq:alt_breeder}) we write as
\[
\mathbf{R}'=(\Omega+E)\Omega^{-1}\mathbf{S}'
\]

\section{Discussion \& Conclusion.}

The breeder's equation is a cornerstone of quantitative genetics and
appears as a fundamental equation in all the important textbooks of
this field \cite{Lynch1998,Falconer1995,Crow2009}. It is widely used
in artificial selection \cite{Lush1943,Hill2010}; its usage in natural
selection was popularized by Lande \cite{LANDE1976} when he formalized
the main idea of phenotypic evolution and it is now commonly used
in many articles based on Lande's work (see for example \cite{Manna2011,Svardal2011,Hansen2011}
). The mathematical foundation of this equation rests upon the hypothesis
that the breeding value is normally distributed. This hypothesis is
plausible for a continuous trait in a population not subject to selection
(see however Appendix/Segregation density function). The normal distribution
of the breeding value is more fragile in populations subjected to
selection on this trait \cite{TURELLI1990}, as the genotype of selected
parents is given by (eq. \ref{eq:pselectedparents}) 
\[
p_{w}(y)=p_{0}(y).W^{\dagger}(y)/\bar{W}
\]
where $W^{\dagger}(y)$ is the genotype fitness function defined by
eq. (\ref{eq:Wtilde}). Even if $p_{0}(y)$ were Gaussian, the very
act of multiplying it by an arbitrary function makes $p_{w}(y)$,
and hence $p_{1}(y)$ non-Gaussian. Therefore after the first round
of selection, the normal distribution hypothesis of parental genotype
cannot be sustained. Turelli and Barton \cite{TURELLI1994} have shown
that for the infinitesimal model, the non-normality may not have large
effects on the predictions of the breeder's equation, but they argued
that when the number of loci is limited the discrepancy can grow much
larger. Of course even $p_{0}(y)$ cannot be assumed to be Gaussian
if different breeds are crossed to constitute the parental generation,
which happens in artificial selection and in natural selection when
gene flow from nearby patches is important. 

The breeding value is not an observable quantity. The fitness or selection
function $W(z)$ is more quantifiable and many authors have considered
a Gaussian selection function. In artificial selection, it dates back
at least to the work of Lush \cite{Lush1943}, p140). In natural selection,
it is used by most authors as a model for stabilizing selection. If
Gaussian selection is used to evolve a population, then the alternative
breeding equation (\ref{eq:alt_breeder}) we derived is more precise
and predictive and rests on more robust mathematical grounds while
retaining the same simplicity of the standard breeder's equation.
Note that the analysis of this article is not restricted to the infinitesimal
model, but applies to all inheritance processes involving purely additive
genetic effects. The alternative breeder's equation generalizes to
selection on multiple traits in a similar way to the standard breeder's
equation and can therefore be incorporated in the ``adaptive landscape''
formalism \cite{Arnold2001} with the same ease. 

In conclusion, we believe that in all cases where Gaussian selection
functions are used to evolve a population, the alternative breeder's
equation we develop above is a useful alternative approach to the
standard method.

\paragraph{Acknowledgment. }

I thank Jarrod Hadfield for discussions and comments on earlier drafts
that helped to improve the manuscript. I am also grateful to M. Vallade,
E. Geissler, O. Rivoire and A. Dawid for careful reading of the manuscript
and fruitful discussions. The author declares to have no conflict
of interest. 

\bibliographystyle{plainnat}
\bibliography{/home/bahram/0Papers/Quant_Genetics/qg}

\newpage{}

\section{Appendix.}

\subsection*{Fourier Transforms and convolutions.\label{sec:FT}}

The Fourier Transform (FT) of a function $f(x)$ is defined here as
\cite{Byron1992} 
\[
\tilde{f}(k)=\mbox{TF}\left[f(x)\right]=\int_{-\infty}^{+\infty}f(x)e^{-ikx}dx
\]
where $i^{2}=-1$. The main properties of FT we use here are (i) Parseval's
theorem
\[
\int_{-\infty}^{+\infty}f^{*}(x)g(x)dx=\frac{1}{2\pi}\int_{-\infty}^{+\infty}\tilde{f}^{*}(k)\tilde{g}(k)dk
\]
where $a^{*}$ stands for the conjugate complex of $a$ ; (ii) the
derivation property
\[
i\frac{d}{dk}\tilde{f}(k)=\int_{-\infty}^{+\infty}xf(x)e^{-ikx}dx=\mbox{TF}\left[xf(x)\right]
\]
(iii) the convolution property
\[
\mbox{FT}\left[(f*g)(x)\right]=\mbox{FT}\left[f(x)\right].\mbox{FT}\left[g(x)\right]=\tilde{f}(k).\tilde{g}(k)
\]
Based on the above properties, and the fact that all the above functions
are real \emph{i.e., }for example\emph{ $W^{*}(z)=W(z)$}, we see
that  relation (\ref{eq:genresponse}) can be written as 
\begin{eqnarray*}
\iint_{\mathbb{R}^{2}}yp_{0}(y)W^{*}(z)f(z-y)dydz & = & \int_{\mathbb{R}}W^{*}(z)\left(yp_{0}*f\right)(z)dz\\
 & = & \frac{i}{2\pi}\int_{\mathbb{R}}\tilde{W}^{*}(k)\tilde{p}'(k)\tilde{f}(k)dk
\end{eqnarray*}
where we have used the fact (i) that $\mbox{FT}\left[yp_{0}(y)\right]=i\tilde{p}'(k)$
; (ii) FT transforms a convolution product into a simple product in
reciprocal space and (iii) Parseval's theorem. 

The same set of rules leads to 
\begin{eqnarray*}
\iint_{\mathbb{R}^{2}}zp_{0}(y)W^{*}(z)f(z-y)dydz & = & \int_{\mathbb{R}}W^{*}(z)z\left(p_{0}*f\right)(z)dz\\
 & = & \frac{i}{2\pi}\int_{\mathbb{R}}\tilde{W}^{*}(k)\frac{d}{dk}\left[\tilde{p}(k)\tilde{f}(k)\right]dk
\end{eqnarray*}

Note that we can exchange the order of integration on $y$ and $z$,
write the first integral as a convolution product on functions of
$z$ and proceed to the second integral by using the Fourier Transform
on $y$. For $R$, we have 
\begin{eqnarray*}
\iint_{\mathbb{R}^{2}}yp_{0}(y)W(z)f(z-y)dydz & = & \int_{\mathbb{R}}p_{0}^{*}(y)y\left(W*f\right)(y)dy\\
 & = & \frac{i}{2\pi}\int_{\mathbb{R}}\tilde{p}^{*}(k)\frac{d}{dk}\left[\tilde{W}(k)\tilde{f}(k)\right]dk
\end{eqnarray*}
and for $S$ we get 

\begin{eqnarray*}
\iint_{\mathbb{R}^{2}}zp_{0}(y)W(z)f(z-y)dydz & = & \int_{\mathbb{R}}p_{0}^{*}(y)\left(zW*f\right)(y)dy\\
 & = & \frac{i}{2\pi}\int_{\mathbb{R}}\tilde{p}^{*}(k)\tilde{W}'(k)\tilde{f}(k)dk
\end{eqnarray*}

The translation property of Fourier Transforms 
\[
\mbox{FT}\left[W_{c}(z-\mu)\right]=e^{-ik\mu}\mbox{FT}\left[W_{c}(z)\right]
\]
was used in the derivation of the functional lags (\ref{eq:Sprime},\ref{eq:Rprime}).

Finally, note that the FT of a Gaussian is a Gaussian : 
\[
\mbox{FT}\left[\frac{1}{\sqrt{2\pi}s}\exp(-x^{2}/(2s^{2})\right]=\exp\left(-s^{2}k^{2}/2\right)
\]

\subsection*{Parent-Offspring regression.\label{sec:PO-regression}}

The derivation of the breeder's equation sometimes uses the parent-offspring
regression coefficient as an intermediate\cite{Lynch1998,Nagylaki1992}.
The linear regression between parent and offspring phenotype however
is based on the same assumption of normal distribution of genotype
and environmental factors. 

The probability density of observing the phenotype $z'$ in the offspring
and $z_{a},z_{b}$ in the parents is 
\begin{eqnarray*}
p(z';z_{a},z_{b}) & = & \iint_{\mathbb{R}^{2}}p(y_{a})f(z_{a}|y_{a})p(y_{b})f(z_{b}|y_{b})\\
 &  & L\left(y_{1}-(y_{a}+y_{b})/2\right)f(z'|y_{1})dy_{1}dy_{a}dy_{b}
\end{eqnarray*}
and the conditional expectation of $z'$ given $z$ is 
\[
E(z'|z_{a},z_{b})=\int_{z'\in I}z'p(z',z)dz'/\int_{z'\in I}p(z',z)dz'=F(z_{a},z_{b})
\]
It is not difficult to check that the function $F(z_{a},z_{b})$ is
a linear function of its argument 
\[
F(z_{a},z_{b})=b(z_{a}+z_{b})/2
\]
if both the genotype and environment factors obey a normal distribution,
in which case, the linearity coefficient is indeed $b=\sigma_{A}^{2}/(\sigma_{A}^{2}+\sigma_{E}^{2})$.
However, \emph{even} if the parental generation follows a normal distribution,
the \emph{selected }parents do not (equation \ref{eq:pselectedparents})
and the use of parent-offspring regression poses even more of a problem
than the direct method.

\subsection*{Segregation density function.\label{sec:Reproduction}}

Let $p_{0}(y)$ be the distribution of breeding value in the parental
generation. In the absence of selection, after recombination-segregation,
the distribution of breeding value in the progeny is 
\begin{equation}
p_{1}(y)=\iint_{\mathbb{R}^{2}}p_{0}(y_{a})p_{0}(y_{b})L\left(y-(y_{a}+y_{b})/2\right)dy_{a}dy_{b}\label{eq:transmissionfunction}
\end{equation}
where the function $L(y)$ is the segregation density function capturing
the inheritance process of the breeding value \cite{KARLIN1979}.
$L(y)$ is a probability density function and in the absence of epistasis
and dominance effect, its average is zero : $\int_{\mathbb{R}}yL(y)dy=0$.
In the infinitesimal model framework, $L(y)$ is a normal distribution
of variance $\sigma_{A}^{2}/2$ . However, any distribution probability
$L(y)$ will lead to a stable, although not necessarily normal, probability
distribution of breeding values after few round of reproduction. Let
us set the origin of the breeding value at its average in the parental
distribution, \emph{i.e.} $\int_{\mathbb{R}}yp_{0}(y)dy=0$. In Fourier
space relation (\ref{eq:transmissionfunction}) is 
\[
\tilde{p}_{1}(k)=\tilde{p}_{0}^{2}(k/2)\tilde{L}(k)
\]
and after $n$ rounds of reproduction, 
\[
\tilde{p}_{n}(k)=\tilde{p}_{0}^{2^{n}}(k/2^{n})\prod_{i=0}^{n-1}\tilde{L}^{2^{i}}(k/2^{i})
\]
As both $p_{0}(y)$ and $L(y)$ are probability distribution functions
of zero mean, we have 
\begin{eqnarray*}
\tilde{p}_{0}(0) & = & \tilde{L}(0)=1\\
\tilde{p}'_{0}(0) & = & \tilde{L'}(0)=0
\end{eqnarray*}
and therefore 
\[
\tilde{p}''_{n}(0)=\frac{1}{2^{n}}\tilde{p}''_{0}(0)+(2-\frac{1}{2^{n-1}})\tilde{L}''(0)
\]
Let $V=\int_{\mathbb{R}}y^{2}L(y)dy$ . We see then that
\[
\mbox{Var}(Y_{n})=\frac{1}{2^{n}}\mbox{Var}(Y_{0})+(2-\frac{1}{2^{n-1}})V
\]
So the variance of the breeding values converges fast to twice the
variance of the segregation density function. The distribution function
$p_{n}(y)$ however converges to a normal distribution only if $L(y)$
is normal.

\subsection*{Individual based numerical simulations.\label{sec:Individual-based-numerical}}

The numerical simulations are performed with the Matlab (Mathwork
inc.) program. $N$ individuals (usually $10^{6}$ ) are generated
and stored in a genotype table $y0$, the genotype $y0(i)$ of individual
$i$ is drawn from a given zero-mean distribution. A table $\xi0$
of the same size is drawn from a normal distribution $\mathcal{N}(0,\sigma_{E})$
and table $z0=y0+\xi0$ is then generated: $u0=[y0,z0]$ constitutes
the parental genotype-phenotype table. For a given fitness function
$W(z)\le1$, a survival table $r$ of size $N$, drawn from a uniform
distribution $\mathcal{U}(0,1)$ is generated. A logical filter selects
elements in $u0$ if $W\left(z0(i)\right)\ge r(i)$. The selected
elements constitute the new table $uw=[yw,zw]$ of size $N1$. A table
$\xi1$ of size $N1$ is drawn again from a normal distribution $\mathcal{N}(0,\sigma_{E})$
and the phenotype of the offspring is computed by $z1=yw+\xi1$. The
various distributions can now be computed from these tables. The selection
differential and the response are computed in the same way, $R=\mbox{mean}(z1)$
and $S=\mbox{mean}(zw)$. 

The above procedure is the core program and is used in other programs,
for example to measure $R$ and $S$ as a function of the selection
function translation. 
\end{document}